\documentclass[conference]{IEEEtran}
\IEEEoverridecommandlockouts

\RequirePackage[utf8]{inputenc}
\RequirePackage[english]{babel}
\RequirePackage{cite}
\RequirePackage{amsmath,amssymb,amsfonts}
\RequirePackage{algorithmic}
\RequirePackage[pscoord]{eso-pic}
\RequirePackage{textcomp}
\RequirePackage{xcolor}
\RequirePackage[a4paper, total={184mm,239mm}]{geometry}
\PassOptionsToPackage{detect-all, per-mode=symbol, range-units=single, range-phrase=~to~}{siunitx}
\RequirePackage{siunitx}
\RequirePackage{nicefrac}
\RequirePackage[hidelinks]{hyperref}
\RequirePackage{orcidlink}
\RequirePackage[noabbrev,capitalise]{cleveref}
\RequirePackage{float}
\RequirePackage[caption=false]{subfig}
\RequirePackage{url}
\RequirePackage{lipsum}
\RequirePackage{placeins}
\RequirePackage{xfrac}
\RequirePackage{multirow}

\RequirePackage{amssymb}
\RequirePackage{pifont}

\RequirePackage{threeparttable}
\RequirePackage{booktabs}
\RequirePackage{colortbl}
\RequirePackage{tabularx}
\RequirePackage{makecell}

\RequirePackage{graphicx}
\RequirePackage{glossaries}
\RequirePackage{xspace}
\RequirePackage{svg}
\usepackage{caption}
\definecolor{ieee-bright-dblue-100}{rgb}{0.0, 0.3828, 0.6055}
\definecolor{ieee-bright-dblue-80}{rgb}{0.0, 0.4883, 0.6797}
\definecolor{ieee-bright-dblue-60}{rgb}{0.3633, 0.6094, 0.7617}
\definecolor{ieee-bright-dblue-40}{rgb}{0.5898, 0.7383, 0.8398}
\definecolor{ieee-bright-dblue-20}{rgb}{0.8906, 0.8984, 0.9219}
\definecolor{ieee-bright-red-100}{rgb}{0.7266, 0.0469, 0.1836}
\definecolor{ieee-bright-red-80}{rgb}{0.832, 0.3164, 0.3281}
\definecolor{ieee-bright-red-60}{rgb}{0.8906, 0.4922, 0.4805}
\definecolor{ieee-bright-red-40}{rgb}{0.9336, 0.6562, 0.6406}
\definecolor{ieee-bright-red-20}{rgb}{0.9688, 0.8203, 0.8125}
\definecolor{ieee-bright-orange-100}{rgb}{0.9961, 0.6367, 0.0}
\definecolor{ieee-bright-orange-80}{rgb}{0.9844, 0.6953, 0.3125}
\definecolor{ieee-bright-orange-60}{rgb}{0.9883, 0.7695, 0.4844}
\definecolor{ieee-bright-orange-40}{rgb}{0.9922, 0.8359, 0.6562}
\definecolor{ieee-bright-orange-20}{rgb}{0.9961, 0.9219, 0.8164}
\definecolor{ieee-bright-yellow-100}{rgb}{0.9961, 0.8164, 0.0}
\definecolor{ieee-bright-yellow-80}{rgb}{0.9961, 0.8477, 0.2148}
\definecolor{ieee-bright-yellow-60}{rgb}{0.9961, 0.875, 0.4492}
\definecolor{ieee-bright-yellow-40}{rgb}{0.9961, 0.9062, 0.6328}
\definecolor{ieee-bright-yellow-20}{rgb}{0.9961, 0.9531, 0.8125}
\definecolor{ieee-bright-lgreen-100}{rgb}{0.4688, 0.7422, 0.125}
\definecolor{ieee-bright-lgreen-80}{rgb}{0.5742, 0.7852, 0.332}
\definecolor{ieee-bright-lgreen-60}{rgb}{0.6875, 0.8398, 0.5039}
\definecolor{ieee-bright-lgreen-40}{rgb}{0.793, 0.8906, 0.6641}
\definecolor{ieee-bright-lgreen-20}{rgb}{0.8945, 0.9414, 0.8281}
\definecolor{ieee-bright-dgreen-100}{rgb}{0.0, 0.5156, 0.2383}
\definecolor{ieee-bright-dgreen-80}{rgb}{0.1641, 0.6055, 0.3867}
\definecolor{ieee-bright-dgreen-60}{rgb}{0.3906, 0.6953, 0.5234}
\definecolor{ieee-bright-dgreen-40}{rgb}{0.6094, 0.8008, 0.6719}
\definecolor{ieee-bright-dgreen-20}{rgb}{0.8047, 0.8945, 0.8359}
\definecolor{ieee-bright-purple-100}{rgb}{0.5938, 0.1133, 0.5898}
\definecolor{ieee-bright-purple-80}{rgb}{0.6992, 0.3281, 0.668}
\definecolor{ieee-bright-purple-60}{rgb}{0.7812, 0.4961, 0.7461}
\definecolor{ieee-bright-purple-40}{rgb}{0.8555, 0.6602, 0.8281}
\definecolor{ieee-bright-purple-20}{rgb}{0.9219, 0.8281, 0.9023}
\definecolor{ieee-bright-lblue-100}{rgb}{0.0, 0.6094, 0.6484}
\definecolor{ieee-bright-lblue-80}{rgb}{0.0, 0.6797, 0.7188}
\definecolor{ieee-bright-lblue-60}{rgb}{0.2109, 0.75, 0.7812}
\definecolor{ieee-bright-lblue-40}{rgb}{0.5469, 0.8242, 0.8438}
\definecolor{ieee-bright-lblue-20}{rgb}{0.7695, 0.918, 0.9219}
\definecolor{ieee-bright-cyan-100}{rgb}{0.0, 0.707, 0.8828}
\definecolor{ieee-bright-cyan-80}{rgb}{0.0, 0.7227, 0.9453}
\definecolor{ieee-bright-cyan-60}{rgb}{0.2656, 0.7812, 0.957}
\definecolor{ieee-bright-cyan-40}{rgb}{0.5547, 0.8438, 0.9688}
\definecolor{ieee-bright-cyan-20}{rgb}{0.7773, 0.9141, 0.9805}
\definecolor{ieee-bright-white-100}{rgb}{0.9961, 0.9961, 0.9961}
\definecolor{ieee-bright-white-80}{rgb}{0.9961, 0.9961, 0.9961}
\definecolor{ieee-bright-white-60}{rgb}{0.9961, 0.9961, 0.9961}
\definecolor{ieee-bright-white-40}{rgb}{0.9961, 0.9961, 0.9961}
\definecolor{ieee-bright-white-20}{rgb}{0.9961, 0.9961, 0.9961}
\definecolor{ieee-dark-red-100}{rgb}{0.5234, 0.1211, 0.2539}
\definecolor{ieee-dark-red-80}{rgb}{0.6445, 0.2812, 0.3828}
\definecolor{ieee-dark-red-60}{rgb}{0.7422, 0.4727, 0.5234}
\definecolor{ieee-dark-red-40}{rgb}{0.832, 0.6445, 0.6758}
\definecolor{ieee-dark-red-20}{rgb}{0.918, 0.8203, 0.832}
\definecolor{ieee-dark-orange-100}{rgb}{0.9062, 0.4648, 0.1328}
\definecolor{ieee-dark-orange-80}{rgb}{0.9648, 0.5664, 0.3164}
\definecolor{ieee-dark-orange-60}{rgb}{0.9766, 0.6758, 0.4805}
\definecolor{ieee-dark-orange-40}{rgb}{0.9844, 0.7773, 0.6523}
\definecolor{ieee-dark-orange-20}{rgb}{0.9922, 0.8789, 0.8125}
\definecolor{ieee-dark-yellow-100}{rgb}{0.9961, 0.7773, 0.1719}
\definecolor{ieee-dark-yellow-80}{rgb}{0.9961, 0.8086, 0.375}
\definecolor{ieee-dark-yellow-60}{rgb}{0.9961, 0.875, 0.4492}
\definecolor{ieee-dark-yellow-40}{rgb}{0.9961, 0.8984, 0.6875}
\definecolor{ieee-dark-yellow-20}{rgb}{0.9961, 0.9453, 0.8438}
\definecolor{ieee-dark-lgreen-100}{rgb}{0.3945, 0.5508, 0.0938}
\definecolor{ieee-dark-lgreen-80}{rgb}{0.5078, 0.6289, 0.293}
\definecolor{ieee-dark-lgreen-60}{rgb}{0.6367, 0.7188, 0.4688}
\definecolor{ieee-dark-lgreen-40}{rgb}{0.7539, 0.8047, 0.6367}
\definecolor{ieee-dark-lgreen-20}{rgb}{0.875, 0.9023, 0.8125}
\definecolor{ieee-dark-dgreen-100}{rgb}{0.0, 0.3867, 0.2539}
\definecolor{ieee-dark-dgreen-80}{rgb}{0.1836, 0.5, 0.3906}
\definecolor{ieee-dark-dgreen-60}{rgb}{0.3984, 0.6172, 0.5273}
\definecolor{ieee-dark-dgreen-40}{rgb}{0.5938, 0.7422, 0.6758}
\definecolor{ieee-dark-dgreen-20}{rgb}{0.793, 0.8711, 0.8359}
\definecolor{ieee-dark-purple-100}{rgb}{0.4648, 0.1445, 0.5117}
\definecolor{ieee-dark-purple-80}{rgb}{0.5898, 0.3242, 0.6016}
\definecolor{ieee-dark-purple-60}{rgb}{0.6914, 0.4883, 0.6953}
\definecolor{ieee-dark-purple-40}{rgb}{0.7969, 0.6523, 0.793}
\definecolor{ieee-dark-purple-20}{rgb}{0.8945, 0.8203, 0.8945}
\definecolor{ieee-dark-cyan-100}{rgb}{0.0, 0.4492, 0.4648}
\definecolor{ieee-dark-cyan-80}{rgb}{0.0, 0.5469, 0.5664}
\definecolor{ieee-dark-cyan-60}{rgb}{0.3047, 0.6602, 0.668}
\definecolor{ieee-dark-cyan-40}{rgb}{0.5586, 0.7695, 0.7734}
\definecolor{ieee-dark-cyan-20}{rgb}{0.7734, 0.8789, 0.8789}
\definecolor{ieee-dark-dblue-100}{rgb}{0.0, 0.1562, 0.332}
\definecolor{ieee-dark-dblue-80}{rgb}{0.1797, 0.3008, 0.4609}
\definecolor{ieee-dark-dblue-60}{rgb}{0.3828, 0.4609, 0.5859}
\definecolor{ieee-dark-dblue-40}{rgb}{0.5781, 0.6289, 0.7188}
\definecolor{ieee-dark-dblue-20}{rgb}{0.7852, 0.8047, 0.8555}
\definecolor{ieee-dark-grey-100}{rgb}{0.457, 0.4688, 0.4805}
\definecolor{ieee-dark-grey-80}{rgb}{0.5625, 0.5625, 0.5742}
\definecolor{ieee-dark-grey-60}{rgb}{0.6641, 0.6641, 0.6758}
\definecolor{ieee-dark-grey-40}{rgb}{0.7734, 0.7695, 0.7773}
\definecolor{ieee-dark-grey-20}{rgb}{0.8789, 0.8828, 0.8828}
\definecolor{ieee-dark-black-100}{rgb}{0.0, 0.0, 0.0}
\definecolor{ieee-dark-black-80}{rgb}{0.3438, 0.3477, 0.3555}
\definecolor{ieee-dark-black-60}{rgb}{0.5, 0.5078, 0.5195}
\definecolor{ieee-dark-black-40}{rgb}{0.6523, 0.6602, 0.6719}
\definecolor{ieee-dark-black-20}{rgb}{0.8164, 0.8242, 0.8281}

\definecolor{light-gray}{gray}{0.75}

\newcommand{\x}{$\times$}

\renewcommand{\subsubsection}[1]{\paragraph*{\textbf{#1}}}
\DeclareSIUnit{\x}{\!\ensuremath{\times}}
\DeclareSIUnit\bit{b}
\DeclareSIUnit\GE{GE}
\DeclareSIUnit\kGE{\kilo\GE}
\DeclareSIUnit\MGE{\mega\GE}
\def\vclic{vCLIC\xspace} % Virtualized CLIC

\renewcommand{\baselinestretch}{0.98}
\begin{document}

\bstctlcite{IEEEexample:BSTcontrol}

% copyright for arXiv
% Disclaimer
%\AddToShipoutPictureBG*{%
%  \AtPageUpperLeft{%
%    \hspace{\paperwidth}%
%    \raisebox{-\baselineskip}{%
%      \makebox[-35pt][r]{\footnotesize{
%        \copyright~2023~IEEE. Personal use of this material is permitted. %
%        Permission from IEEE must be obtained for all other uses, in any current or future media, including
%      }}
%}}}%

%\AddToShipoutPictureBG*{%
%  \AtPageUpperLeft{%
%    \hspace{\paperwidth}%
%   \raisebox{-2\baselineskip}{%
%      \makebox[-37pt][r]{\footnotesize{
%        reprinting/republishing this material for advertising or promotional purposes, creating new collective works, for resale or redistribution to servers or lists, or
%      }}
%}}}%

%\AddToShipoutPictureBG*{%
%  \AtPageUpperLeft{%
%    \hspace{\paperwidth}%
%    \raisebox{-3\baselineskip}{%
%      \makebox[-185pt][r]{\footnotesize{
%       reuse of any copyrighted component of this work in other works.
%      }}
%}}}%

%%%%%%%%%%%%%%%%%%%%%%
%%   VERSIONING     %%
%%%%%%%%%%%%%%%%%%%%%%

% version
\ifx\showrevision\undefined
    \newcommand{\todo}[1]{{#1}}
    \newcommand{\ph}[1]{{#1}}
\else
    \newcommand{\todo}[1]{{\textcolor{ieee-bright-red-80}{#1}}\PackageWarning{TODO:}{#1!}}
    \newcommand{\ph}[1]{{\textcolor{light-gray}{#1}}\PackageWarning{PH:}{#1!}}
    \AddToShipoutPictureFG{%
        \put(%
            8mm,%
            \paperheight-1.5cm%
            ){\vtop{{\null}\makebox[0pt][c]{%
                \rotatebox[origin=c]{90}{%
                    \huge\textcolor{ieee-bright-red-80!75}{\reviewpass}%
                }%
            }}%
        }%
    }
    \AddToShipoutPictureFG{%
        \put(%
            \paperwidth-6mm,%
            \paperheight-1.5cm%
            ){\vtop{{\null}\makebox[0pt][c]{%
                \rotatebox[origin=c]{90}{%
                    \huge\textcolor{ieee-bright-red-80!30}{ETH Zurich - Unpublished - Confidential - Draft - Copyright 2024}%
                }%
            }}%
        }%
    }
\fi

% Include acronyms
% General
\newacronym{sota}{SOTA}{state-of-the-art}
\newacronym{fpga}{FPGA}{field programmable gate array}
\newacronym{asic}{ASIC}{application-specific integrated circuit}
\newacronym{fub}{FUB}{functional unit block}
\newacronym{vv}{V\&V}{validation and verification}
\newacronym{gpp}{GPP}{general purpose processor}
\newacronym{cots}{COTS}{commercial off-the-shelf}

% Domain terminology
\newacronym{hpc}{HPC}{high performance computing}
\newacronym{ml}{ML}{machine learning}
\newacronym{isa}{ISA}{instruction set architecture}
\newacronym{fp}{FP}{floating-point}
\newacronym{dl}{DL}{deep learning}
\newacronym{la}{LA}{linear algebra}
\newacronym{ip}{IP}{intellectual property}
\newacronym[firstplural=systems-on-chip (SoCs)]{soc}{SoC}{system-on-chip}
\newacronym{mpsoc}{MPSoC}{multi-processor system-on-chip}
\newacronym[firstplural=networks-on-chip (NoCs)]{noc}{NoC}{network-on-chip}
\newacronym{hw}{HW}{hardware}
\newacronym{sw}{SW}{software}
\newacronym{swapc}{SWaP-C}{space, weight, power, and cost}
\newacronym{mcp}{MCP}{multi-core processor}
\newacronym{rr}{RR}{round-robin}

% Methodology
\newacronym{mac}{MAC}{multiply-accumulate}
\newacronym{fem}{FEM}{finite element analysis}
\newacronym{simd}{SIMD}{single-instruction, multiple-data}
\newacronym{rtl}{RTL}{register transfer level}
\newacronym{dlt}{DLT}{data layout transform}

% Hardware units
\newacronym{fifo}{FIFO}{first in, first out}
\newacronym{fu}{FU}{functional unit}
\newacronym{alu}{ALU}{arithmetic logic unit}
\newacronym{fpu}{FPU}{floating-point unit}
\newacronym{ssr}{SSR}{stream semantic register}
\newacronym{issr}{ISSR}{indirection stream semantic register}
\newacronym{tcdm}{TCDM}{tightly-coupled data memory}
\newacronym{dma}{DMA}{direct memory access}
\newacronym{sm}{SM}{streaming multiprocessor}
\newacronym{vlsu}{VLSU}{vector load-store unit}
\newacronym{dsa}{DSA}{domain-specific accelerator}
\newacronym{ha}{HA}{hardware accelerator}
\newacronym{fsm}{FSM}{finite state machine}
\newacronym{llc}{LLC}{last-level cache}
\newacronym{d2d}{D2D}{die-to-die}
\newacronym{dram}{DRAM}{dynamic random access memory}
\newacronym{tid}{TID}{transaction ID}
\newacronym{spm}{SPM}{scratchpad memory}

% Software
\newacronym{os}{OS}{operating system}
\newacronym{rtos}{RTOS}{real-time operating system}
\newacronym{gpos}{GPOS}{general-purpose operating system}

% Technologies
\newacronym{axi4}{AXI4}{Advanced eXtensible Interface 4}
\newacronym{amba}{AMBA}{Advanced Microcontroller Bus Architecture}
\newacronym{sram}{SRAM}{static random-access memory}

% Real-time
\newacronym{wcet}{WCET}{worst-case execution time}
\newacronym{rtunit}{REALM unit}{real-time regulation and traffic monitoring unit}
\newacronym{mtunit}{M\&R unit}{monitoring and regulation unit}
\newacronym{cps}{CPS}{cyber-physical system}
\newacronym{crtes}{CRTES}{critical real-time embedded system}
\newacronym{heicps}{He-iCPS}{heterogeneous integrated cyber-physical system}
\newacronym{ecu}{ECU}{electronic control unit}
\newacronym{mcs}{MCS}{mixed criticality system}
\newacronym{ima}{IMA}{integrated modular avionics}
\newacronym{adas}{ADAS}{advanced driver assistance system}
\newacronym{axirealm}{AXI-REALM}{AXI real-time regulation and traffic monitoring}
\newacronym{mpam}{MPAM}{memory system resource partitioning and monitoring}
\newacronym{dos}{DoS}{denial of service}
\newacronym{hwrot}{HWRoT}{hardware root of trust}
\newacronym{clic}{CLIC}{Core-Local Interrupt Controller}
\newacronym{plic}{PLIC}{Platform-Level Interrupt Controller}
\newacronym{msi}{MSI}{message-signaled interrupt}
\newacronym{aia}{AIA}{advanced interrupt architecture}
\newacronym{vm}{VM}{virtual machine}
\newacronym{vs}{VS}{virtual supervisor}
\newacronym{hs}{HS}{hypervisor supervisor}
\newacronym{vmm}{VMM}{Virtual Machine Monitor}
\newacronym{pmp}{PMP}{physical memory protection}
\newacronym{vsid}{VSID}{virtual supervisor ID}
\newacronym{csr}{CSR}{control and status register}
% \newacronym{ic}{IC}{integrated circuit}
\newacronym{pc}{PC}{program counter}
\newacronym{sph}{SPH}{static partitioning hypervisor}
\newacronym{dph}{DPH}{dynamic partitioning hypervisor}
\newacronym{isr}{ISR}{interrupt service routine}
\newacronym{abi}{ABI}{application binary interface}
\newacronym{eabi}{EABI}{embedded application binary interface}
\newacronym{hart}{HART}{hardware thread}
\newacronym{ooo}{OoO}{order of magnitude}
\newacronym{risc}{RISC}{reduced instruction set computing}
\newacronym{shv}{SHV}{selective hardware vectoring}
\newacronym{imsic}{IMSIC}{Incoming MSI Controller}
\newacronym[firstplural=interrupt controllers (ICs)]{ic}{IC}{interrupt controller}

%%%%%%%%%%%%%%%%%%%%%%
%%   FRONT MATTER   %%
%%%%%%%%%%%%%%%%%%%%%%

\title{vCLIC: Towards Fast Interrupt Handling in Virtualized RISC-V Mixed-criticality Systems
\thanks{This work was supported in part by the TRISTAN project (101095947) that received funding from the HORIZON CHIPS-JU programme.}}

% balasr: maybe context switching doesn't make sense
% ale: I changed

\ifx\blind\undefined
    \author{
        % \IEEEauthorblockN{%
        % Enrico Zelioli\orcidlink{0009-0005-4899-1047}\IEEEauthorrefmark{1}, %
        % Alessandro Ottaviano\orcidlink{0009-0000-9924-3536}\IEEEauthorrefmark{1}, %
        % Robert Balas\orcidlink{0000-0002-7231-9315}\IEEEauthorrefmark{1}, %
        % Nils Wistoff\orcidlink{0000-0002-8683-8060}\IEEEauthorrefmark{1}, % 
        % Angelo Garofalo\orcidlink{0000-0002-7495-6895}\IEEEauthorrefmark{1}\IEEEauthorrefmark{2}, %
        % Luca Benini\orcidlink{0000-0001-8068-3806}\IEEEauthorrefmark{1}\IEEEauthorrefmark{2}%
        % }
        \IEEEauthorblockN{%
        Enrico Zelioli\IEEEauthorrefmark{1}, %
        Alessandro Ottaviano\IEEEauthorrefmark{1}, %
        Robert Balas\IEEEauthorrefmark{1}, %
        Nils Wistoff\IEEEauthorrefmark{1}, % 
        Angelo Garofalo\IEEEauthorrefmark{1}\IEEEauthorrefmark{2}, %
        Luca Benini\IEEEauthorrefmark{1}\IEEEauthorrefmark{2}%
        }
        \IEEEauthorblockA{
            \IEEEauthorrefmark{1}IIS, ETH Zurich, Switzerland;
            \IEEEauthorrefmark{2}DEI, University of Bologna, Italy \\
            \{ezelioli, aottaviano, balasr, nwistoff, agarofalo, lbenini\}@iis.ee.ethz.ch
        }
    }

% \author{\IEEEauthorblockN{Enrico Zelioli}
%     \IEEEauthorblockA{\textit{Integrated Systems Laboratory} \\
%     \textit{ETH Zurich}\\
%     Zurich, Switzerland \\
%     ezelioli@iis.ee.ethz.ch}
%     \and
%     \IEEEauthorblockN{Alessandro Ottaviano}
%     \IEEEauthorblockA{\textit{Integrated Systems Laboratory} \\
%     \textit{ETH Zurich}\\
%     Zurich, Switzerland \\
%     aottaviano@iis.ee.ethz.ch}
%     \and
%     \IEEEauthorblockN{Robert Balas}
%     \IEEEauthorblockA{\textit{Integrated Systems Laboratory} \\
%     \textit{ETH Zurich}\\
%     Zurich, Switzerland \\
%     balasr@iis.ee.ethz.ch}
%     \and
%     \IEEEauthorblockN{Nils Wistoff}
%     \IEEEauthorblockA{\textit{Integrated Systems Laboratory} \\
%     \textit{ETH Zurich}\\
%     Zurich, Switzerland \\
%     nwistoff@iis.ee.ethz.ch}
%     \and
%     \IEEEauthorblockN{Angelo Garofalo}
%     \IEEEauthorblockA{\textit{Integrated Systems Laboratory} \\
%     \textit{ETH Zurich}\\
%     Zurich, Switzerland \\
%     agarofalo@iis.ee.ethz.ch}
%     \and
%     \IEEEauthorblockN{Luca Benini}
%     \IEEEauthorblockA{\textit{Integrated Systems Laboratory} \\
%     \textit{ETH Zurich}\\
%     Zurich, Switzerland \\
%     lbenini@iis.ee.ethz.ch}
% }
\else
    \author{%
            \vspace{1.1cm} %
            \textit{Authors omitted for blind review}
            % \vspace{1.1cm} %
            }
\fi

\maketitle

% required words count: <= 300 words for ICCD
% current word count (update if changed): 220
\begin{abstract}

The widespread diffusion of compute-intensive edge-AI workloads and the stringent demands of modern autonomous systems require advanced heterogeneous embedded architectures. Such architectures must support high-performance and reliable execution of parallel tasks with different levels of criticality. Hardware-assisted virtualization is crucial for isolating applications concurrently executing these tasks under real-time constraints, but interrupt virtualization poses challenges in ensuring transparency to virtual guests while maintaining real-time system features, such as interrupt vectoring, nesting, and tail-chaining. Despite its rapid advancement to address virtualization needs for mixed-criticality systems, the RISC-V ecosystem still lacks interrupt controllers with integrated virtualization and real-time features, currently relying on non-deterministic, bus-mediated message-signaled interrupts (MSIs) for virtualization.

To overcome this limitation, we present the design, implementation, and in-system assessment of vCLIC, a virtualization extension to the RISC-V CLIC fast interrupt controller. Our approach achieves 20$\times$ interrupt latency speed-up over the software emulation required for handling non-virtualization-aware systems, reduces response latency by 15\% compared to existing MSI-based approaches, and is free from interference from the system bus, at an area cost of just 8kGE when synthesized in an advanced 16nm FinFet technology.
\end{abstract}


%%%%%%%%%%%%%%%%%%
%%   CONTENT   %%%
%%%%%%%%%%%%%%%%%%

\section{Introduction}
\label{sec:intro}

Applications like autonomous driving, navigation, and robotics require safety, security, and energy efficiency for edge-AI tasks~\cite{AUTOMOTIVE_PREDICTABLE}, driving the need for heterogeneous \glspl*{mcs}. As \gls*{sw} complexity grows, so does the integration of \glspl*{gpos} in embedded systems. However, they can only partially support the specialized real-time features of \glspl*{rtos}, which prioritize worst-case execution time. Virtualization has emerged as a key technology to enable the simultaneous operation of \glspl*{gpos} and \glspl*{rtos} on the same \gls*{hw} platform~\cite{HEISER_VIRT}. In safety-critical systems, real-time guarantees are essential, especially regarding low and deterministic interrupt latency. For such systems, \gls*{hw}-assisted virtualization of resources --- e.g., \glspl*{ic} and executing cores --- are vital to minimize hypervisor intervention, reducing response latency and jitter.
% \gls*{cots}

Commercial off-the-shelf general-purpose embedded systems often feature diverse \gls*{ic} architectures to allocate and sequence interrupts based on specific application needs. 
For instance, Arm's generic \gls*{ic} (GIC) is versatile and can manage wired and \glspl*{msi} at different design scales. 
Albeit GICv3 and GICv4 support partial hardware-assisted virtualization --- for example, direct injection of interrupts to \glspl*{vm} is not supported ---, they are not designed to handle interrupt nesting, and their real-time capabilities are limited to a particular channel for fast and critical interrupts.
%  (FIQ)
In contrast, the Nested Vectored Interrupt Controller (NVIC) is specifically designed for real-time operations in a non-virtualized setting on small embedded devices, with interrupt latencies and context switching times as low as 12 and 96 clock cycles, respectively~\cite{yiu_cortex_reference_13}. 
% removed citation from cortex_m4_manual
%
Similarly, RISC-V introduced the \gls*{clic}, optimized for real-time applications, showcasing features like interrupt nesting and tail-chaining. However, there is no support for virtualization of the \gls*{clic} in the most recent specifications~\cite{clic}. On the other end, the RISC-V \gls*{aia} caters to larger, high-performance virtualized systems, but without support for time-critical interrupt handling~\cite{aia}. Notably, efforts to extend the \gls*{aia} with real-time and predictability features have been made, mitigating the bus interference effects typical of \gls*{msi}-based systems~\cite{IEAIA_MINHO}. Although improving predictability of virtual \glspl*{msi}, AIA-based solutions still lack support for real-time interrupt handling features. Sa et al. enhanced the RISC-V \gls*{plic} to support virtualization~\cite{VPLIC}. While significantly improving the best-case interrupt latency over alternative software emulation approaches, the \gls*{plic} does not support any of the aforementioned real-time features.

Therefore, existing \gls*{ic} solutions do not entirely meet the diverse requirements of real-time, compact, virtualized embedded \glspl*{mcs} at once. Focusing on the growing, open RISC-V ecosystem as an ideal domain for design space exploration and \gls*{hw}/\gls*{sw} co-design, we fill this gap by presenting {\vclic}, a novel, lightweight virtualization extension for the \gls*{clic}.
%Similarly, RISC-V introduced the \gls*{clic}, optimized for real-time applications showcasing features like interrupt nesting and tail-chaining, and the \gls*{aia}, which caters to larger, high-performance virtualized systems without support for time-critical interrupt handling~\cite{aia}.
%
%We posit that existing \gls*{ic} solutions do not entirely meet the diverse requirements of compact, virtualized embedded \glspl*{mcs}.
%Focusing on the growing RISC-V ecosystem as an ideal domain for design space exploration and \gls*{hw}/\gls*{sw} co-design, we advocate for the RISC-V \gls*{clic} as a promising solution for this market segment (\cref{sec:back}).
%
%There is no mention of a virtualized variant of the \gls*{clic} in the most recent specifications~\cite{clic}.
%To fill this gap, in this work we present {\vclic}, a novel, lightweight virtualization extension to the \gls*{clic}. 
%
We evaluate its impact on system size and demonstrate its benefits on a virtualized, embedded RISC-V platform built around the CVA6 application-class core~\cite{CVA6}. The {\vclic} synthesizable \gls*{hw} description is freely accessible and
\ifx\blind\undefined
     open-source.\footnote{\texttt{\url{https://github.com/pulp-platform/clic}}}
\else
     open-source.\footnote{\texttt{URL omitted for blind review}}
\fi
% 
% Contributions
\subsubsection{Contribution}
The contributions of our work include:
% \vspace{-1pt}
\begin{enumerate}

    \item \textbf{{\vclic} extension} We design and implement an interrupt virtualization extension of the RISC-V CLIC and the application-class CVA6 core (\cref{sec:arch}). {\vclic} enables direct injection of interrupts to guest \glspl*{vm} and prioritization of real-time, safety-critical \glspl*{vm}.
    
    \item \textbf{In-system implementation assessment} We integrate {\vclic} in Cheshire, a Linux-capable RISC-V \gls*{soc}~\cite{10163410}, and characterize it in a \SI{16}{\nano\metre} FinFet technology node. 
    The virtualization extension in its minimal configuration incurs an area overhead of just \SI{8}{\kilo\GE} at iso-frequency with the vanilla CLIC design and a total overhead of just \SI{1}{\%} at the \gls*{soc} level (\cref{subsec:eval:impl}). 

    \item \textbf{In-system functional assessment} We demonstrate the beneficial effects of low interrupt latency and jitter of {\vclic}. Our approach achieves 20x speed-up over the software emulation method required by non-virtualized systems (\cref{subsec:eval:func}). Besides reducing response latency by 15\% compared to existing \gls*{msi}-based virtualization strategies (RISC-V \gls*{aia}), {\vclic} is free from interference with the system bus. 
    
\end{enumerate}

% Useful from Krste: https://github.com/riscv/riscv-fast-interrupt/issues/92

%The AIA spec is being developed to be the standard support for interrupts for standard hypervisors. Possible hypervisor directions for fast interrupts for embedded include: 
% 1) a minimal CLIC extension that adds HS mode as another privilege layer in the stack, so the hypervisor could optionally receive interrupts and optionally enable/hide interrupts from lower privilege modes. This would allow the 2-stage translation and other features of the hypervisor to be used in a CLIC-based system, but would not support sending interrupts directly to descheduled guest OS contexts (have to route via hypervisor interrupt). 
% 2) support multiple descheduled CLIC contexts to which interrupts could be sent directly while they were sleeping, which would be a much bigger project. 
% 3) add an extension to AIA (instead of working on CLIC) to somehow reduce interrupt latency using the standard hypervisor stack.
% \input{text/Background}
\section{Architecture}
\label{sec:arch}

% \begin{figure}[t]
%     \centering
%     \subfloat[Cheshire platform diagram. Per-core vCLIC is highlighted.]{\label{fig:cheshire:system}\includegraphics[width=\columnwidth]{fig/cheshire.pdf}}\\
%     \subfloat[CLIC and CVA6 interface. Only relevant parts of the core are shown.]{\label{fig:cheshire:zoomin}\includegraphics[width=\columnwidth]{fig/vCLIC-CVA6.pdf}}%
%     \vspace{.1cm}
%     \caption{Cheshire (a) and its interrupt subsystem (b).}
%     \label{fig:cheshire}
% \end{figure}

% \begin{figure}[t]
%     \centering
%     \includegraphics[width=\columnwidth]{fig/interrupt_flow.pdf}
%     % \includesvg[width=\columnwidth]{fig/interrupt_flow.svg}
%     \caption{The interrupt handling process with a virtualized CLIC (figure above) and with a vanilla CLIC (figure below).}
%     \label{fig:flow}
% \end{figure}

%Our design builds on top of the CLIC specification by adding a configurable \gls*{vs} extension to support direct injection of interrupts to \gls*{vm} guests. 
In virtualized RISC-V systems, the {\vclic} extension allows a hypervisor to delegate individual interrupts to \glspl*{vm}. The delegation reduces interrupt latency by eliminating the hypervisor-mediated \gls*{sw} emulation of the \gls*{ic}. The {\vclic} is integrated into the Cheshire platform~\cite{10163410}, briefly described in~\cref{subsec:arch:cheshire}. The {\vclic} design is split into two modular extensions: (i) \texttt{VSCLIC}, which enables the delegation of interrupt lines to \glspl*{vm}, and (ii) \texttt{VSPRIO}, which allows the hypervisor to prioritize interrupts depending on the \gls*{vm} they are assigned to, as discussed in~\cref{subsec:arch:vclic}. Furthermore, the enhancements required in the core (in this work, CVA6) connected to the {\vclic} are described in~\cref{subsec:arch:cva6}. \Cref{fig:cheshire:zoomin} details the {\vclic}-CVA6 interface, highlighting the novel \gls*{hw} features compared to vanilla \gls*{clic}.

\subsection{Cheshire Platform}\label{subsec:arch:cheshire}

Cheshire~\cite{10163410} is a minimal system built around the 64-bit, application-class CVA6 RISC-V processor~\cite{CVA6}. The platform is highly configurable, supporting seamless integration of domain-specific accelerators or other general-purpose engines with diverse criticality levels via an AXI4 fully connected crossbar, as shown in \cref{fig:cheshire}. 
%\Cref{fig:cheshire} illustrates Cheshire's high-level block diagram. 
%It shows how incoming interrupts from peripheral devices are masked and distributed to each core via a programmable interrupt router and then locally managed by the corresponding {\vclic}.
In a multi-core setup, each {\vclic} instance is tightly coupled to each core, ensuring independence from the system bus and thereby eliminating interference at the interconnect level. % --- a critical feature in \glspl*{mcs}, where shared resources often become performance bottlenecks and complicate real-time analysis~\cite{benz2023axirealm}.

\subsection{{\vclic} extension}\label{subsec:arch:vclic}

% \vspace{0.1cm}

\subsubsection{VSCLIC}\label{subsubsec:arch:vsclic}

\begin{figure}[t]
    \centering
    \subfloat[Cheshire platform diagram. Per-core vCLIC is highlighted.]{\label{fig:cheshire:system}\includegraphics[width=0.95\columnwidth]{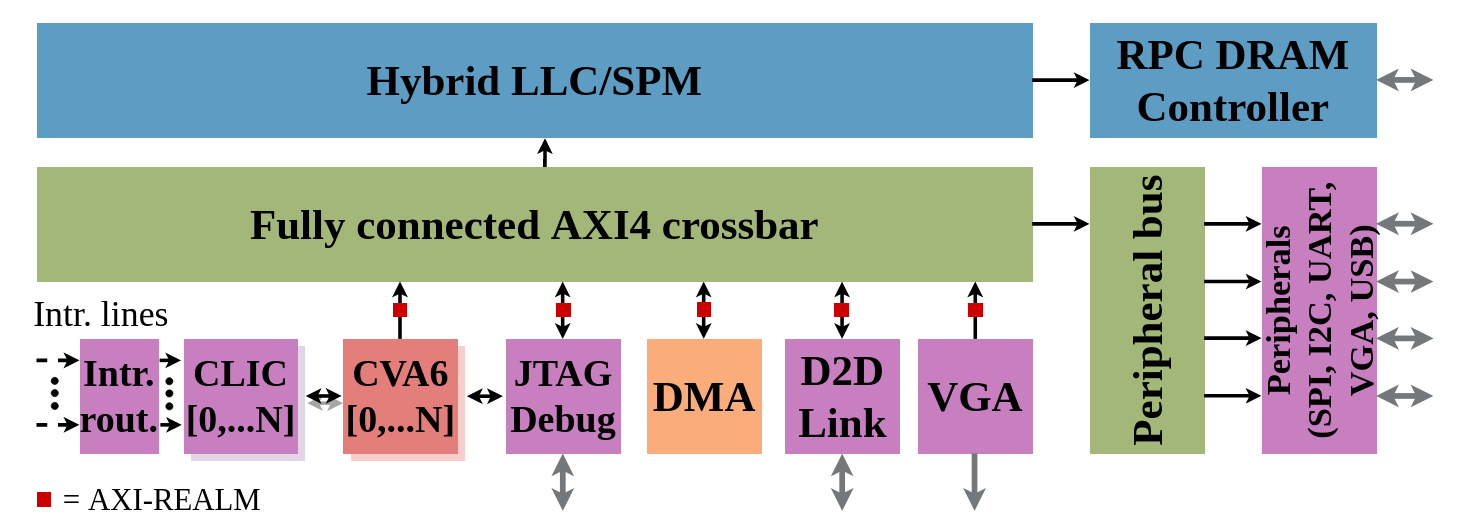}}\\
    \subfloat[CLIC and CVA6 interface. Only relevant parts of the core are shown.]{\label{fig:cheshire:zoomin}\includegraphics[width=0.95\columnwidth]{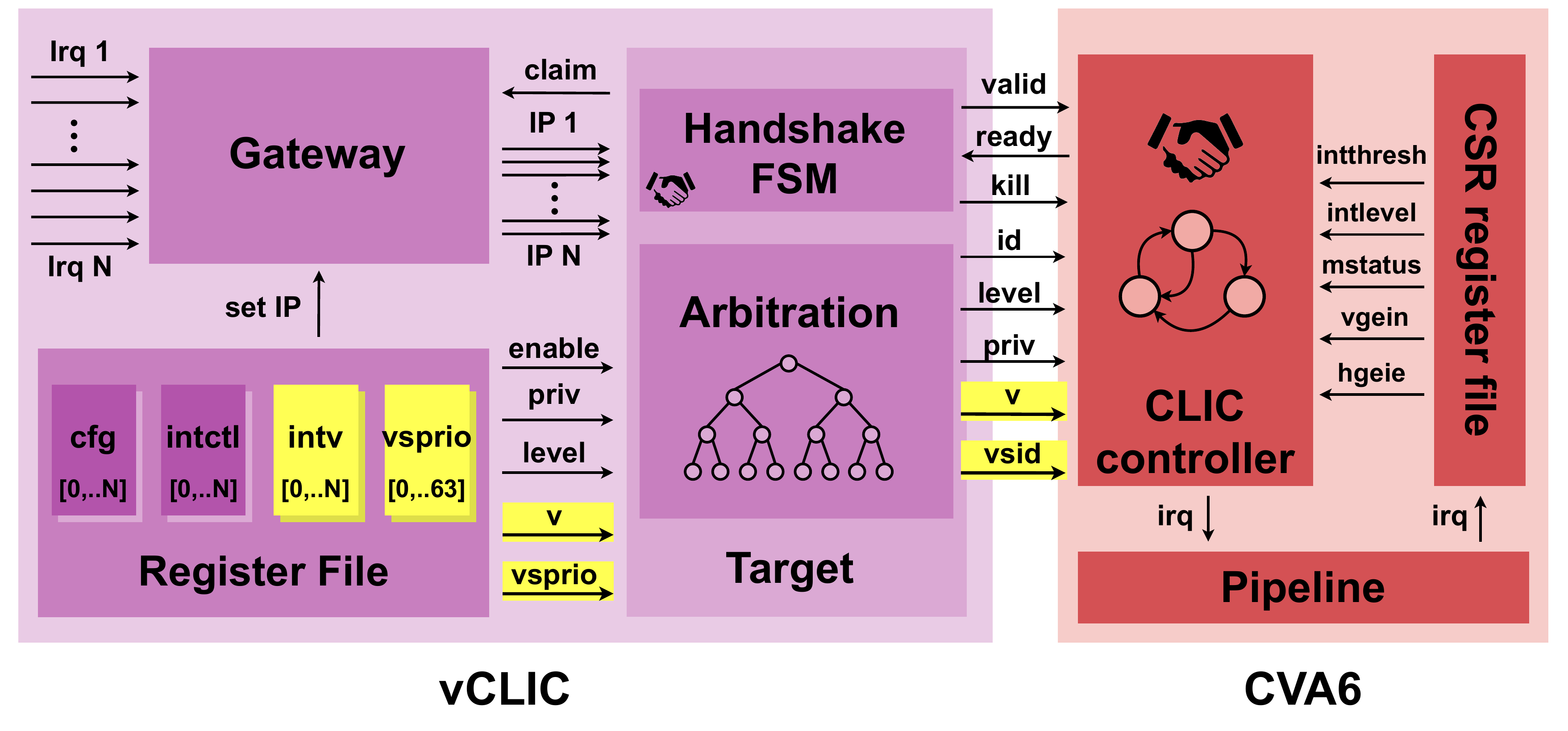}}
    \vspace{.1cm}
    \caption{Cheshire (a) and its interrupt subsystem (b).}
    \label{fig:cheshire}
\end{figure}

The \texttt{VSCLIC} extension allows to delegate interrupt lines to \gls*{vs} mode. It introduces a new set of memory-mapped configuration registers, namely \texttt{clicintv}, that are under the control of the hypervisor (\cref{fig:cheshire:zoomin}). For each interrupt line \emph{i}, \texttt{clicintv[\emph{i}]} is a \SI{1}{\byte} register comprising (i) a virtualization bit (\emph{v} field), which adds a new orthogonal privilege mode to the privilege mode interrupt attribute, and (ii) a virtual supervisor ID (\emph{vsid} field), which holds the \gls*{vsid} of the \gls*{vm} to which the interrupt is delegated.

%Since each interrupt line can only be assigned to one \emph{interrupt target} (i.e. a software component identified by privilege mode or guest ID) at a time, the 
{\vclic} multiplexes the virtual configuration registers of each target on a single set of physical registers, minimizing the area overhead on the register file. 
%The configuration of interrupt lines delegated to higher-privilege modes appears to be hardwired to zero to lower-privileged software. 
The privilege of software accessing {\vclic} configuration registers is determined by the address range used to access them by partitioning the {\vclic} configuration address space. To guarantee isolation, each \gls*{vm} must be granted access to only one of these address regions (i.e., the one associated with its \gls*{vsid}). Such isolation can be enforced by privileged software by means of \gls*{pmp} units or --- on systems supporting two-stage address translation --- by virtual memory.
 
% \vspace{0.1cm}

\subsubsection{VSPRIO}\label{subsubsec:arch:vsprio}

The {\vclic} supports up to 64 simultaneous active guests, identified by a \gls*{vsid} stored in the \emph{VGEIN} field of the \emph{hstatus} \gls*{csr}, as per the \mbox{RISC-V} \texttt{H} extension. This allows to statically delegate interrupt lines of pass-through peripherals to \glspl*{vm}, eliminating the need of saving and restoring the \gls*{ic} configuration upon \gls*{vm} context switch, thereby improving the worst-case interrupt latency. 

Nevertheless, to ensure that the real-time properties of interrupts delegated to safety-critical guests are not affected, the \texttt{VSPRIO} extension allows the hypervisor to prioritize virtual interrupts based on the associated \glspl*{vm}. The extension enlarges the interrupt priority space by adding a set of privileged configuration registers that hold a per-\gls*{vm} priority value. This enables dynamic prioritization of real-time and safety-critical \glspl*{vm}, allowing the deployment of systems with more tightly bound worst-case interrupt latencies. 

% \begin{figure}[t]
%     \centering
%     \subfloat[Cheshire platform diagram. Per-core vCLIC is highlighted.]{\label{fig:cheshire:system}\includegraphics[width=\columnwidth]{fig/cheshire.pdf}}\\
%     \subfloat[CLIC and CVA6 interface. Only relevant parts of the core are shown.]{\label{fig:cheshire:zoomin}\includegraphics[width=\columnwidth]{fig/vCLIC-CVA6.pdf}}
%     \vspace{.1cm}
%     \caption{Cheshire (a) and its interrupt subsystem (b).}
%     \label{fig:cheshire}
% \end{figure}

\subsection{CVA6 extension}\label{subsec:arch:cva6}

We extend CVA6's interrupt handling logic to behave as follows when a \gls*{vm} is running: (i) if the incoming interrupt belongs to a higher privilege level, the core traps to that privilege level; (ii) if the incoming interrupt is directed to the currently running \gls*{vm} (i.e., the interrupt's and running \gls*{vm}'s \gls*{vsid} match), the interrupt is serviced by the running \gls*{vm}; (iii) if the incoming interrupt is directed to a different \gls*{vm}, the core traps to hypervisor mode if and only if the target \gls*{vm} has higher priority. Remarkably, item (iii) is crucial for the hypervisor to regain control of the core in case of high-priority events that must be handled by the dedicated \gls*{vm} with minimal latency.

% Furthermore, the {\vclic} extension introduces new \glspl*{csr} and extends the functionality of others. 
Furthermore, the virtualization extension requires the following \glspl*{csr} to be added: (i) \texttt{VSINTTHRESH}, the interrupt \emph{level} threshold (managing interrupt nesting in \gls*{clic}-based systems), (ii) \texttt{VSTVT}, the guest physical address of \gls*{vs} trap vector table (in \gls*{shv} mode), and (iii) \texttt{VSNXTI} for interrupt tail-chaining. In CLIC mode, the \texttt{vsie}, \texttt{vsip}, and \texttt{vsideleg} appear hardwired to zero, as their functionality is embedded in the \gls*{ic} itself. Finally, {\vclic} supports the original CLIC \gls*{shv} extension, allowing fine-grained management of virtual interrupts vectoring.

% removed citation from ~\cite{buttazzo2011hard, liu1973scheduling}

% \input{text/archi_2}
\section{Evaluation}
\label{sec:eval}
% We evaluate the performance of {\vclic} by measuring interrupt latency and jitter (as defined in~\cref{subsec:back:term}) in a virtualized environment and assessing the effects of micro-architectural state and resource contention on such metrics. 
% %
% Furthermore, we assess the area overhead of our solution compared to the baseline \gls*{clic}. Finally, we compare our results with those of related state-of-the-art solutions.

\subsection{Framework}\label{subsec:eval:frame}

To carry out fast and cycle-accurate latency measurements, we integrate {\vclic} into the Cheshire \gls*{soc} shown in~\cref{fig:cheshire} and emulate the system on a Digilent Genesys 2 FPGA board. On FPGA, Cheshire is configured with one CVA6 core and {\vclic} pair, 64 input interrupt lines, \SI{32}{\kibi\byte} of data and instruction cache, \SI{128}{\kibi\byte} of last-level cache, and is constrained to run at \SI{50}{\mega\hertz}. As a \gls*{sw} stack, we use two Hypervisors: the Bao \gls*{sph}~\cite{Martins2020BaoAL} and the Xvisor \gls*{dph}~\cite{XVISOR}, both open-source. 
Bao allows pinning a virtual guest to a physical core, improving the overall system determinism. We use it to assess the interrupt latency and jitter of {\vclic} in a built-in, predictable \gls*{sw} setting. Xvisor, in contrast, allows multiple \gls*{vm} guests to be assigned to a single physical core. We use Xvisor to assess interrupt prioritization and preemption across \gls*{vm} guests. 
% ({\vclic}'s \texttt{VSPRIO} extension). 
% We evaluate our implementation by comparing the interrupt latency of the FreeRTOS \gls*{rtos} with different \glspl*{ic}. To estimate the effect of resource contention in \glspl*{dph}, we run Linux and FreeRTOS on top of XVisor and measure the worst-case interrupt latency upon context-switch events. 

\subsection{Functional}\label{subsec:eval:func}

\begin{figure}[t]
    \centering
    \includegraphics[width=\columnwidth]{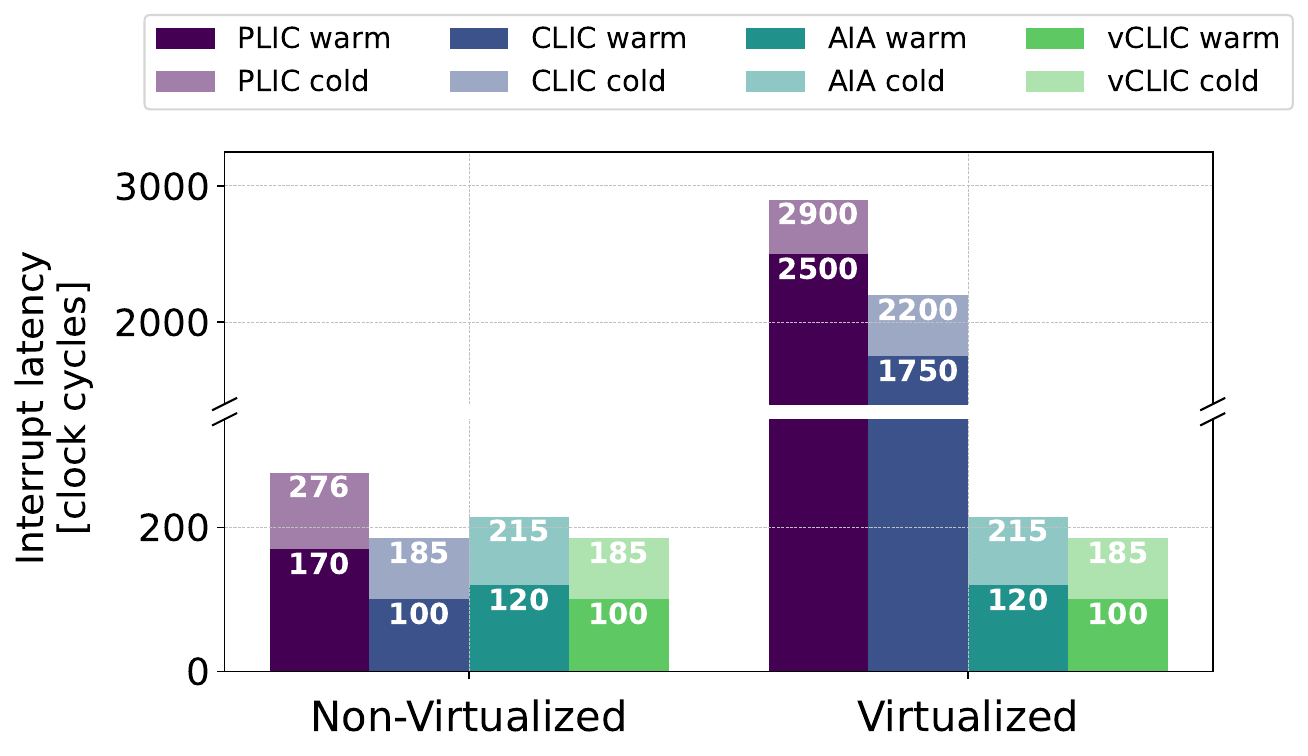}
    % \caption{Interrupt latency comparison for \textit{non-virtualized}, or bare-metal (left-hand side), and \textit{virtualized}, running FreeRTOS \gls*{vm} on top of Bao (right-hand side). 
    % % Each bar represents a different interrupt subsystem integrated in Cheshire and deployed on FPGA.
    % }
    \caption{Interrupt latency comparison for \textit{bare-metal} (left), and \textit{virtualized} with FreeRTOS on top of Bao (right) scenarios. 
    % Each bar represents a different interrupt subsystem integrated in Cheshire and deployed on FPGA.
    }
    \label{fig:latency}
\end{figure}

% \begin{figure}[t]
%     \centering
%     % \includegraphics[width=\columnwidth]{fig/latency_breakdown.pdf}
%     \includegraphics[width=\columnwidth]{fig/variability.pdf}
%     \caption{Interrupt latency breakdown and impact of caches and TLBs architectural state. CVA6 executes a bare-metal guest \gls*{vm} on top of the Bao \gls*{vmm}.}
%     \label{fig:latency_breakdown}
% \end{figure}

% \begin{figure}[t]
%     \centering
%     \includesvg[width=\columnwidth]{fig/context_switch.svg}
%     \caption{Impact of {\vclic}'s \texttt{VSPRIO} extension. 
%     We show the \gls*{clic} context switching overhead with 64 interrupts per \gls*{vm}. We use Linux and FreeRTOS as guest \glspl*{vm} under XVisor. The target platform is Cheshire deployed on FPGA.
%     %We show that \glspl*{vm} interrupt prioritization can reduce the time to completion of a time-critical task when multiple guests are executing on the same physical resource.
%     }
%     \label{fig:contextsw}
% \end{figure}

\subsubsection{Interrupt latency}

Interrupt latency includes both \gls*{hw} and \gls*{sw} contributions. We include \gls*{sw} contributions affected by the core and \gls*{ic} design in the interrupt latency measurement to ensure a fair comparison. These are, for instance, the time to save the interrupt context and to identify the interrupt source.
% decode the interrupt (e.g., by reading an interrupt ID from a dedicated register).

Following this definition, we evaluate the interrupt latency of external interrupt sources with various interrupt subsystems: (i) vanilla \mbox{RISC-V} PLIC implementation~\cite{plic}; (ii) vanilla \mbox{RISC-V} CLIC implementation~\cite{clic}; (iii) vanilla AIA implementation, i.e. one A-PLIC and one \gls*{imsic}~\cite{costaopen}; and (iv) vCLIC proposed in this work. 
% Each subsystem is integrated into Cheshire and deployed on FPGA, as described in~\cref{subsec:eval:frame}. 

On the \gls*{sw} side, CVA6 executes in two settings: non-virtualized, or bare-metal, and virtualized, using FreeRTOS \gls*{vm} on top of the Bao monitor. We collect the interrupt latencies by running each experiment 100 times, assuming a warm micro-architectural state (i.e., we do not flush the caches and TLBs between iterations). This setup represents a lower bound on the expected interrupt latency, which can be affected by several factors, as discussed in the following sections.

\Cref{fig:latency} reports the best (minimum) and worst (maximum) interrupt latency of a single external interrupt source (e.g., a programmable timer) for each scenario. We observe a significant increase of about 20$\times$ in the interrupt latency of configurations (i) and (ii) without \gls*{hw}-assisted interrupt virtualization. The increase is due to the overhead of the \gls*{ic} \gls*{sw} emulation that the hypervisor must perform in the absence of \gls*{hw} support. The PLIC emulation results in a larger overhead due to the extra memory access required by the PLIC to read the interrupt ID that must be emulated by the hypervisor. Virtualization does not affect interrupt latency for both the \gls*{aia} and the {\vclic} configurations. The {\vclic} achieves a lower interrupt latency than \gls*{aia} due to the use of wired interrupts instead of interconnect-dependent \glspl*{msi}. 

\subsubsection{Interrupt jitter}

We identify two main contributions to the interrupt jitter: (i) instruction and data caches and (ii) TLBs. 
We assume two states for each component, i.e., \textit{warm} (recently accessed) and \textit{cold} (not recently accessed). We assess the impact of such components by measuring the interrupt latency of a custom bare-metal guest with virtual memory support under varying initial states of each element. On average, we observe that the interrupt latency --- independently from the \gls*{ic} --- is about 8$\times$ higher when caches are cold. As expected, cache misses only impact the \gls*{sw} contributions to the interrupt latency, such as the context save and restore and the interrupt decoding operations.  Conversely, TLB misses caused by cold TLBs result in a moderate \SI{5}{\percent} average slowdown. This happens because, for simple interrupt handlers, the entire \gls*{isr} fits within a single page.
% To enforce the reset of a component to a cold state, we use dedicated RISC-V instructions, i.e., \texttt{fence.i} \cite{RISCV_I} for the instruction cache and \texttt{sfence.vma} for the TLBs.

% To assess the impact of TLBs and cache hierarchy on interrupt latency, we use a custom bare-metal guest with virtual memory support and conduct measurements under varying initial states of each component, resulting in the four scenarios illustrated in \cref{fig:latency_breakdown}. The figure delineates a breakdown of the overall interrupt latency into six primary phases, from core interrupt acknowledgment to return to program execution. While the first four phases directly contribute to the interrupt latency, the remaining (context restore and return to program execution) contribute to the total interrupt handling time.

% \begin{figure}[t]
%     \centering
%     \includegraphics[width=\columnwidth]{fig/interrupt_latency.pdf}
%     % \caption{Interrupt latency comparison for \textit{non-virtualized}, or bare-metal (left-hand side), and \textit{virtualized}, running FreeRTOS \gls*{vm} on top of Bao (right-hand side). 
%     % % Each bar represents a different interrupt subsystem integrated in Cheshire and deployed on FPGA.
%     % }
%     \caption{Interrupt latency comparison for \textit{bare-metal} (left), and \textit{virtualized} with FreeRTOS on top of Bao (right) scenarios. 
%     % Each bar represents a different interrupt subsystem integrated in Cheshire and deployed on FPGA.
%     }
%     \label{fig:latency}
% \end{figure}

\subsection{System resources contention}\label{subsec:eval:contention}

% \Cref{subsec:eval:func} focuses on a single interrupt source targeting a \gls*{vm} running on top of a \gls*{sph}. We do not capture the worst-case interrupt latency by limiting the analysis to such a use case. \glspl*{mcs} typically have strict requirements for predictability and worst-case execution times to guarantee real-time performance and responsiveness to critical events. In this section, we discuss other possible scenarios that impact the worst-case interrupt latency.
To assess the worst-case interrupt latency, we expand our experiments to consider \gls*{hw} resource contention. We identify three main reasons for increased interrupt latency due to resource contention: (i) interrupt arbitration, (ii) bus contention, and (iii) \gls*{vm} interference.

\subsubsection{Interrupt arbitration}
% We refer to \textit{interrupt contention} as the scenario in which multiple interrupt requests are raised simultaneously. 
% Nearly every \gls*{ic} supports some interrupt prioritization. 
The {\vclic} prioritizes interrupts based on the assigned privilege mode, level, and priority, as well as based on a configurable \gls*{vm} priority (\texttt{VSPRIO} extension). However, this only affects the ordering of interrupts before they are serviced. Once an interrupt request is taken, higher-priority incoming interrupts may be delayed depending on the \gls*{ic} and \gls*{sw} implementation. {\vclic} minimizes the impact of interrupt contention through interrupt nesting~\cite{clic,CV32RT}, and extends this feature to virtual interrupts, allowing efficient implementations of preemptible \glspl*{isr}.

\subsubsection{Bus contention}
Existing virtualized systems are commonly based on \glspl*{msi}. \glspl*{msi} are mostly used in server-class computing systems, where predictability and real-time features are not critical requirements. As shown by Marques~et~al.~\cite{IEAIA_MINHO}, some system interconnect topologies can increase the average interrupt latency by up to 7$\times$ due to bus contention for \gls*{aia}-based systems, along with a significant rise (about \SI{750}{\nano\second} at \SI{50}{\mega\hertz}) in the response jitter over time. Unlike \gls*{msi}-based systems, we highlight that our proposed {\vclic} is entirely free from bus interference, regardless of the system interconnect topology, as it relies on core-local wired interrupt lines.

\subsubsection{Inter-\gls*{vm} interference}

In systems relying on \glspl*{dph}, interrupt latencies can vary substantially over time, depending on the resource allocation status at the time of the interrupt. To evaluate such scenarios, we run XVisor with Linux and FreeRTOS as guests. We evaluate the worst-case scenario -- a critical interrupt directed to FreeRTOS arrives while Linux is running -- by measuring the time required by the system to preempt the Linux execution in favor of FreeRTOS. On Cheshire, XVisor takes approximately $35000$ clock cycles to perform a \gls*{vm} context switch. As discussed in section \ref{subsubsec:arch:vsprio}, if the \texttt{VSPRIO} extension is disabled, the hypervisor must also save the configuration of interrupts delegated to the \gls*{vm} being switched out and set the configuration for the next \gls*{vm}. The performance degradation due to the additional interrupt context associated with each \gls*{vm} is proportional to the amount of interrupts delegated to the \gls*{vm}. The \gls*{clic} context switch requires up to $10000$ clock cycles for \glspl*{vm} with 64 virtual interrupts, increasing the overall context switch time by almost $30\%$.
Conversely, the context switch overhead is more tolerable if few interrupts are delegated to a \gls*{vm}, down to about $1250$ clock cycles in case of a single virtual interrupt per \gls*{vm}.

\subsection{Implementation}\label{subsec:eval:impl}

% \begin{figure}[t]
%     \centering
%     \includegraphics[width=0.85\columnwidth]{fig/cheshire_area_breakdown_hbar.pdf}
%     \caption{Cheshire area breakdown, 64 wired input interrupts, synthesis in Intel16 FinFet. {\vclic} is instantiated with \texttt{VSPRIO} extension enabled and 1 priority bit (minimal configuration).}
%     \label{fig:cheshire-footprint}
% \end{figure}

\begin{figure}[t]
    \centering
    \includegraphics[width=0.9\columnwidth]{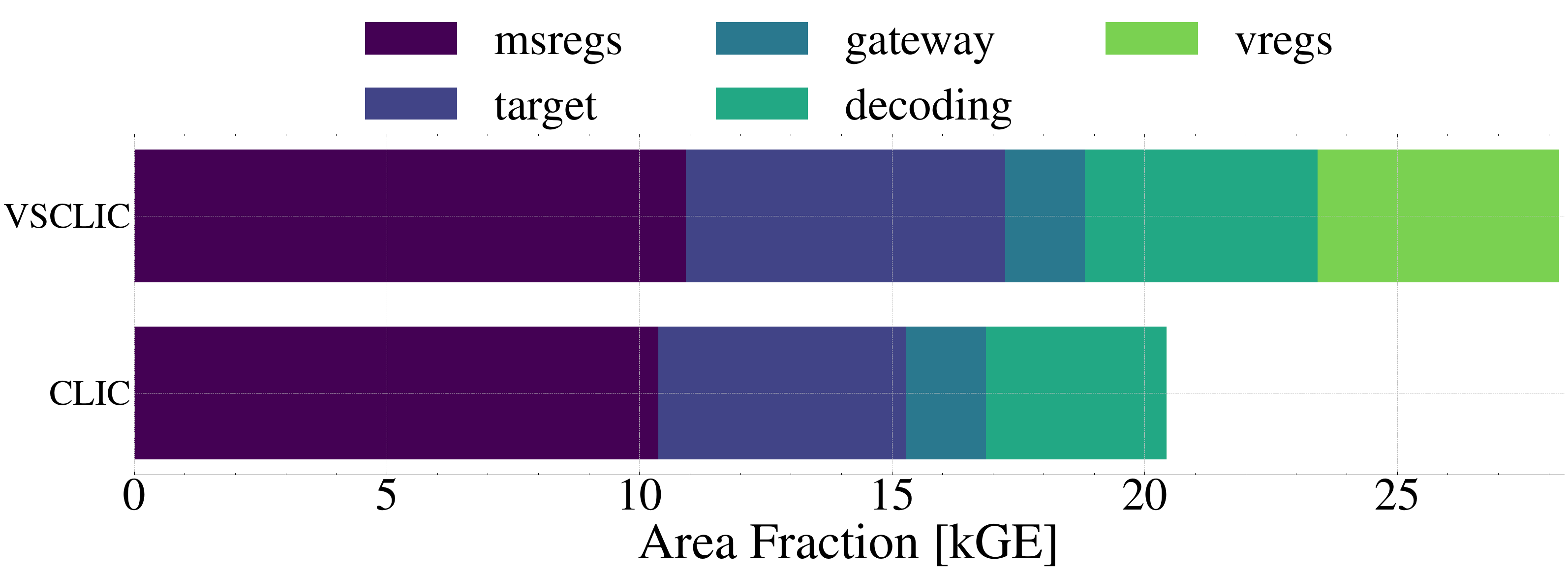}
    \caption{\texttt{VSCLIC} area breakdown compared to vanilla CLIC. 
    % Note the additional memory-mapped registers for virtualization (\texttt{vregs}) and the increased size of the arbitration binary tree (\texttt{target}).
    }
    \label{fig:vsclic-footprint}
\end{figure}

\begin{figure}[t]
    \centering
    \includegraphics[width=0.9\columnwidth]{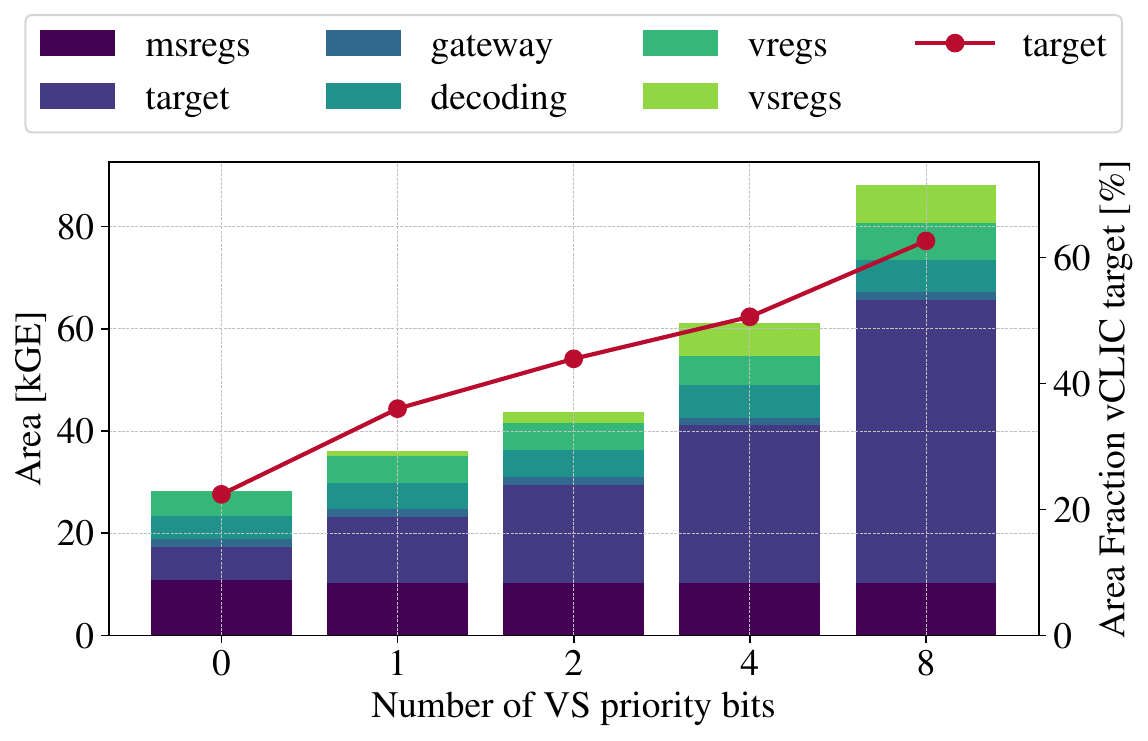}
    \caption{\texttt{VSPRIO} area impact on {\vclic} with the number of priority bits. Case 0 indicates VSCLIC-only implementation.}
    \label{fig:vsprio-footprint}
\end{figure}

We synthesize {\vclic} as part of the Cheshire platform, as described in~\cref{subsec:eval:frame}. We use Synopsys Design Compiler 2022.03, targeting Intel 16 FinFet technology at \SI{1.1}{\giga\hertz}, SS corner, and \SI{125}{\celsius}.
One gate equivalent (GE) for this technology equals \SI{0.233}{\micro\meter\squared}.
Our extension does not affect the critical path of the design.

\Cref{fig:vsclic-footprint} shows the area overhead introduced by the \texttt{VSCLIC} extension compared to vanilla \gls*{clic}, totaling about \SI{8}{\kilo\GE}. Most of the additional resources are due to the increased memory-mapped registers. \gls*{clic}'s binary tree size to propagate the highest level and priority interrupt is equally affected with roughly \SI{27}{\percent} increase. We evaluate the impact of the \texttt{VSPRIO} extension based on the number of priority levels, as shown in~\cref{fig:vsprio-footprint}. The arbitration tree size grows almost linearly with the number of implemented priority bits. 
% We observe that, albeit conducting a detailed space exploration of area variation, a realistic number of priority bits lies in the range 2-4, meaning 4 to 16 priority levels.

Finally, we complete the in-system assessment by measuring the area of Cheshire and the relative impact of {\vclic}. Accounting for the memory macros footprint, {\vclic} occupies merely \SI{1}{\percent} of the system.
Compared to platform-wide centralized approaches, this result justifies a per-core, distributed solution for interrupt management in a virtualized scenario.

\section{Conclusion}
\label{sec:concl}

% To meet the stringent real-time requirements of virtualized embedded \glspl*{mcs} and accommodate the increasing complexity of their software stack, interrupt virtualization becomes essential. In this work, we propose {\vclic}, a virtualization extension to the RISC-V's CLIC fast interrupt controller. We detail its design, implementation, and in-system evaluation. Our approach achieves a 20$\times$ reduction in interrupt latency compared to the \gls*{sw} emulation used in non-virtualized systems, achieving \SI{15}{\percent} faster interrupt response with respect to \gls*{msi}-based approaches,
% %minimizes response latency relative to existing \gls*{msi}-based approaches, 
% and avoids interference with the system bus. When synthesized in a state-of-the-art 16nm FinFet technology node, the minimal configuration of {\vclic} adds only 8kGE of additional logic compared to the non-virtualized CLIC while not affecting the target frequency of the original design, making {\vclic} a low-overhead solution for virtualized real-time embedded \glspl*{mcs}.

To meet the real-time requirements of virtualized embedded \glspl*{mcs} and manage increasing software complexity, interrupt virtualization is crucial. This work introduces {\vclic}, a virtualization extension to the RISC-V CLIC fast interrupt controller. {\vclic} reduces interrupt latency by 20X compared to non-virtualized systems, offers a \SI{15}{\percent} faster response than \gls*{msi}-based approaches, and avoids system bus interference. Synthesized in a 16nm FinFet technology, {\vclic} adds only 8kGE of logic without affecting the target frequency, providing a low-overhead solution for real-time embedded \glspl*{mcs}.

% ack
% \section*{Acknowledgments}
% \label{sec:ack}
% \ifx\blind\undefined
%     %Funding
%     This work was supported in part by the TRISTAN project (101095947) that received funding from the HORIZON CHIPS-JU programme.
% \else
%     \textit{Acknowledgments omitted for blind review.}
% \fi

%%%%%%%%%%%%%%%%%%%%%%%
%%%   BACK MATTER   %%%
%%%%%%%%%%%%%%%%%%%%%%%

% Make sure biblio is never stretched
\renewcommand{\baselinestretch}{1.0}

% TODO: remove
%\newpage

\bibliographystyle{IEEEtran}
\bibliography{IEEEabrv,main}

\end{document}